\documentclass[aps,pre,twocolumn,floatfix,showpacs]{revtex4}

\usepackage{graphicx}
\usepackage{amsmath}
\usepackage{amssymb}

\begin{document}
\title{Kinetic Friction and Atomistic Instabilities in
  Boundary-Lubricated Systems}

\author{Martin Aichele} \affiliation{Institut Charles Sadron, 6 rue
  Boussingault, 67083 Strasbourg, France} \affiliation{Institut f\"ur
  Physik, Universit\"at Mainz, 55099 Mainz, Germany} \author{Martin H.
  M\"user} \affiliation{Department of Applied Mathematics, University
  of Western Ontario, London, Ontario N6A 5B7, Canada} \date{\today}

\begin{abstract}
  The contribution of sliding-induced, atomic-scale instabilities to
  the kinetic friction force is investigated by molecular dynamics.
  For this purpose, we derive a relationship between the kinetic
  friction force $F_{\rm k}$ and the non-equilibrium velocity
  distribution $P(v)$ of the lubricant particles.  $P(v)$ typically
  shows exponential tails, which cannot be described in terms of an
  effective temperature.  It is investigated which parameters control
  the existence of instabilities and how they affect $P(v)$ and hence
  $F_{\rm k}$.  The effects of the interfaces' dimensionality,
  lubricant coverage, and internal degrees of freedom of lubricant
  particles on $F_{\rm k}$ are studied explicitly. Among other results
  we find that the kinetic friction between commensurate surfaces is
  much more susceptible to changes in $(i)$ lubricant coverage, $(ii)$
  sliding velocity, and $(iii)$ bond length of lubricant molecules
  than incommensurate surfaces.
\end{abstract}



\maketitle

\section{Introduction}
\label{sec:intro}
The every-day phenomenon friction is of great practical and economical
importance, which is one of the motivations to improve our
understanding of tribological processes~\cite{persson98,bhushan99}.
Friction between two solids differs from that between a solid and a
fluid in that both static and kinetic friction appear finite, while
the force between a solid and a fluid vanishes linearly with sliding
velocity $v_0$ at small $v_0$.  Static friction $F_{\rm s}$ is the
externally applied force necessary to initiate relative sliding motion
between two solids, whereas kinetic friction $F_{\rm k}$ is the force
needed to maintain the sliding motion.  Phenomenological friction
laws, which date back to \mbox{da~Vinci}, \mbox{Amontons}, and
\mbox{Coulomb}~\cite{dowson79}, often provide a good description on
the macroscopic scale.

The microscopic origins of kinetic friction are still a matter of
debate, even though it has long been recognized that kinetic friction
must be due to dynamical
instabilities~\cite{brillouin09,caroli98epjb}.  While there can be
many different processes leading to instabilities, they all have in
common that potential energy is converted abruptly into kinetic energy
and ultimately lost as heat~\cite{muser03acp}.  Although instabilities
can occur on many different time and length scales, there has been an
enhanced interest in identifying those that occur on atomic scales.
This quest is not only motivated by the miniaturization of technical
devices down to the nanometer scale, but also by the desire to better
understand macroscopic friction.  The understanding of single-asperity
contacts is needed as basis for the full description of macroscopic
friction, where the bulk-mediated coupling between contacts gives rise
to additional effects.

Load-bearing, simple-asperity contacts are often in the order of
microns.  According to Hertzian contact mechanics and generalizations
thereof, the pressure is rather constant in the contact with the
exception of the areas close to the circumference, where pressure
gradients are large.  In the centre of the contact, most of the
lubricant is squeezed out.  One may assume that these
boundary-lubricated areas often account for most of the energy
dissipation when two solids are slid against each other, unless the
solids are very compliant, in which case elastic instabilities may
also contribute a significant amount to the net dissipation.  If wear
was the main source of friction, material would have to rub off from
the surfaces much faster than observed
experimentally~\cite{gnecco02prl}.  Hydrodynamic lubrication would
likewise result in values for friction orders of magnitude too small,
if it were assumed to be the dominant dissipation process.

While the crucial role of surfactants for friction has certainly been
recognized, relatively little attention has been paid to characterize
dynamical instabilities in boundary lubricants. Most of the work on
instabilities leading to friction is devoted to elastic processes,
which are most simply described in the Prandtl-Tomlinson (PT)
model~\cite{prandtl28,tomlinson29}.  In the PT model, an atom is
pulled over a substrate by a spring that moves at constant velocity
$v_0$. If the spring stiffness is below a certain critical value, the
atom's instantaneous velocity can exceed $v_0$ by many orders of
magnitude, see i.e. the discussion in
Refs.~\onlinecite{caroli98epjb,muser03acp}.  This process results in
non-vanishing $F_{\rm k}$ in the limit of zero $v_0$ as long as
thermal fluctuations are absent.  There is, however, a crucial
difference between instabilities in boundary lubricants and
instabilities occurring in elastic manifolds that are modelled in
terms of the PT model and related approaches such as the
Frenkel-Kontorova model~\cite{frenkel38,braun98}.  In boundary
lubricants, atoms are only weakly connected to each other and to the
confining walls. As a consequence, bond breaking can occur, whereas in
elastic models, bonds are treated as unbreakable.  This seemingly
subtle difference leads to different tribological behavior.

Two different avenues have been pursued in the recent past to study
dynamics in boundary lubricants and its consequences for tribological
properties. One is a minimalist approach, in which one single
lubricant atom embedded between two shearing plates is
considered~\cite{rozman96prl,rozman96pre}. In the following, we will
refer to this approach as the impurity limit.  The other avenue
incorporates a large ensemble of lubricant
atoms~\cite{thompson90sci,thompson90pra}.  This approach can
eventually include surface curvature and elastic deformation of the
surfaces making it possible to study what effect the interplay of
surface curvature and elastic deformations have on dry or
boundary-lubricated friction~\cite{wenning01,persson02jcp}.

In this paper, we intend to analyze what features of simplistic models
appear robust as the level of complexity in the description of the
boundary lubricant is increased. Since kinetic friction is intimately
connected with instabilities, we focus on the analysis of
instabilities.  In a precedent paper by one of us~\cite{muser02prl},
it was found that the existence of instabilities in the impurity limit
and as a consequence the friction-velocity relationship $F_{\rm
  k}(v_0)$ depends on the `details' of the model. For instance, it was
found that for 1D, commensurate interfaces, the sign of the first
higher harmonic in the lubricant-wall potential determines: (a)
whether or not the athermal kinetic friction remains finite in the
zero-velocity limit, and (b) the exponent $\beta$ that describes the
finite-velocity corrections by
\begin{equation}
F_{\rm k}(v) - F_{\rm k}(0) \, \propto \, v_0^{\beta}.
\label{eq:athermal_fric}
\end{equation}
Note that Eq.~(\ref{eq:athermal_fric}) changes its form when thermal
noise is included into the treatment, i.e., it becomes linear at small
velocities~\cite{muser02prl}. Depending on the ratio of the relevant
energies and temperature, thermal effects may be negligible down to
very small values of $v_0$.

While Ref.~\onlinecite{muser02prl} is mainly focused on the impurity
limit, we intend to extend the analysis in a systematic fashion to
less idealized situations. For example, instead of simple spherical
impurities, dimers and hexamers (6-mers) will be studied.  Moreover
direct interactions between lubricant particles will be included and
the effects of increasing coverage will be discussed.  The central
assumption of our analysis is the existence of instabilities or `pops'
of certain degrees of freedom. A pop is a sudden, seemingly erratic
motion of a particle (or a collective degree of freedom) characterized
by a velocity much larger than the associated thermal velocity or the
drift velocity of the atom.  Pops heat the lubricant or alternatively
they couple directly to the confining solid walls, i.e., by inducing
phonons in the walls.  They will eventually induce more dramatic
effects such as generation of dislocations or abrade the surfaces.
However, as argued above, these extreme processes are rare and hence
presumably they are not responsible for the main part of the energy
dissipation.  This is the motivation to concentrate on the energy
transfer to the phonon bath that is due to elementary process in the
lubricant.  The underlying idea of the presented approach can be
described as follows. Sliding-induced instabilities make the velocity
probability distribution (PD) of the lubricant atoms deviate from the
thermal equilibrium PD.  This alters the balance of energy flow from
and to the lubricant. The energy missing in this balance is provided
by the external driving device.

In this paper, we will develop a simple kinetic theory that connects
the energy dissipation with the velocity PD
(Section~\ref{sec:theory}). After discussing the numerical techniques
in Sec.~\ref{sec:methods}, we will apply the theory to models of
boundary lubrication of various complexity. This will include both the
impurity limit, which is discussed in Section~\ref{sec:impurity}, and
more complex situations that include interaction between lubricant
atoms (Section~\ref{sec:beyond}).  Section~\ref{sec:conclusions}
contains the conclusions.

\section{Theory}
\label{sec:theory}

\subsection{General Comments}
The most fundamental assumption in this paper is that the interaction
between the lubricant atom $i$ and the confining wall can be
decomposed into one conservative part $V_{\rm w}({\bf r}_i)$ and one
non-conservative term consisting of a damping force plus thermal
noise.  $V_{\rm w}({\bf r}_i)$ depends only on the difference between
the position ${\bf r}_i$ and the positions of top wall ${\bf r}_{\rm
  t}$ and bottom wall ${\bf r}_{\rm b}$. It can be written as
\begin{equation}
V_{\rm w}({\bf r}_i) = V_{\rm b}({\bf r}_i - {\bf r}_{\rm b}) + 
                 V_{\rm t}({\bf r}_i - {\bf r}_{\rm t}),
\end{equation}
where depending on the model under consideration, the vectors ${\bf
  r}$ can be one-, two-, or three-dimensional.  Unless otherwise
noted, the relative motion of the walls is imposed externally, i.e.,
by constant separation (or constant load) and constant relative
velocity $v_0 {\bf e}_x = (\dot{{\bf r}}_{\rm t} - \dot{{\bf r}}_{\rm
  b})$ of the walls parallel to the sliding direction indicated by the
unit vector ${\bf e}_x$. We assume the normal pressure variations to
be small, which means that the coupling to each individual confining
(crystalline) wall is periodic parallel to the interface, i.e., it is
periodic in the $xy$-plane.

In the theoretical part of our treatment, we assume that the {\it
  non-conservative} force ${\bf F}^{\rm (nc)}_{{\rm t},i}$ that a wall
(here the top wall) exerts on the lubricant atom $i$ consists of a
simple, viscous damping term $-\gamma_{\rm t} (\dot{{\bf r}}_i -
\dot{{\bf r}}_{\rm t})$ plus a random force ${\bf \Gamma}_{\rm t}(t)$,
thus
\begin{equation}
{\bf F}^{{\rm (nc)}}_{{\rm t},i} = - m_i \gamma_{\rm t} \left(\dot{{\bf r}}_i - \dot{{\bf r}}_{\rm t}\right) 
+ {\bf \Gamma}_{\rm t}(t),
\label{eq:thermostat}
\end{equation}
where the random force ${\bf \Gamma}(t)$ is chosen such that detailed
balance is obeyed when the external stress is absent. Thus the usual
$\delta$ correlation of the random forces is assumed,
namely~\cite{risken84}
\begin{equation}
\langle {\bf \Gamma}_{\rm t}(t) {\bf \Gamma}_{\rm t}(t') \rangle = 2D\gamma_{\rm t} m_i k_{\rm B} T \delta(t-t'),
\label{eq:gamma_cor}
\end{equation}
where $D$ is the physical dimension, $k_{\rm B} T$ is the thermal
energy, and $m_i$ the mass of lubricant atom $i$.  Random forces plus
damping term $-m_i \gamma_{\rm t} {\bf v}$ (we dub the sum {\it
  thermostat}) mimic the interactions with phonons and/or other
excitations, which are not treated explicitly.  Typically, the time
scales associated with these excitations are short compared to the
motion of a particle from one minimum to another, which justifies the
assumption of \mbox{$\delta$~correlated} random forces for our
purposes.  Of course, damping can and will be different normal and
parallel to the interface. However, this detail does not have any
significant consequences for the conclusions presented in this paper.
Similarly, the explicit treatment of internal elastic deformations
does not alter the major conclusions either.

We will now be concerned with the derivation of a formal equation for
the friction force. In any steady-state of the system, the average
force on the upper wall (or the lower wall) must be zero.  If the time
average was different from zero, the upper wall would be accelerated
in contradiction to the steady-state assumption, as pointed out for
instance, by Thompson and Robbins~\cite{thompson90pra}. The net force
on the upper wall consists of three contributions: The externally
applied force ${\bf F}_{\rm ext}$, the conservative force between
lubricant and wall, and the non-conservative force done by the
thermostat.  The external force ${\bf F}_{\rm ext}$ does the work
$W_{\rm ext}$ on the upper wall given by
\begin{equation}
W_{\rm ext} = \int d {\bf r}_{\rm t} {\bf F}_{\rm ext}.
\label{eq:w_ext}
\end{equation}
The kinetic friction force $F_{\rm k}$ follows from that expression as
it equals the work done on the system by external forces divided by
the distance moved.

Since the conservative potential is assumed to be periodic, it cannot
do any net work on a steady-state system and we may not consider it in
our energy balance.  This implies that the work must be done on the
thermostat. The power dissipated into the damping term is proportional
to $\gamma m (\dot{{\bf r}}-\dot{{\bf r}}_{\rm t})^2$, however, parts
of that contribution can be provided by the stochastic random force
${\bf \Gamma}(t)$.  Hence, if we want to account only for the power
$P_{\rm ext}$ that is dissipated into the damping term due to the {\it
  externally} applied force, we have to integrate over the velocity
distribution $P(v)$ but we have to subtract the contribution that is
due to the random force. The latter contribution is very difficult to
calculate.  We assume that this heat flow from the random force into
the impurity system is identical to that in thermal equilibrium, in
which the equilibrium (Maxwell Boltzmann) distribution $P_{\rm eq}(v)$
applies. This yields
\begin{equation}
P_{\rm ext} = N_{\rm fl} \gamma m \int_0^\infty v^2
\left\{ P(v) - P_{\rm eq}(v) \right\} dv,
\label{eq:dissipation}
\end{equation}
where $N_{\rm fl}$ is the number of lubricant atoms, $v$ is the
velocity of a particle relative to the center-of-mass motion of the
upper wall.  The net external driving force (or in other words the
kinetic friction force $F_{\rm k}$) can now be associated with
\begin{equation}
F_{\rm k} = P_{\rm ext} / v_0.
\label{eq:f_k}
\end{equation}

We want to emphasize that Eqs.~(\ref{eq:w_ext}) through (\ref{eq:f_k})
allow one to calculate friction forces under more general conditions
than those of our particular model, for instance, if the thermostat
only acts on the atoms in the outermost layers of the walls as e.g.\ 
employed in Ref.~\cite{braun01}. The approach can also be extended in
a straightforward manner if generalized forms of the thermostat are
employed such as in dissipative particle dynamics~\cite{espanol95} or
if the thermostat is based on a Mori Zwanzig
formalism~\cite{mori65,zwanzig65}.  The main limitation of
Eq.~(\ref{eq:dissipation}) in the present context is that effects due
to heating of the walls are not included.  Again, a minor modification
would allow one to include heating of the walls into the presented
framework as well.  However, as we will mainly focus on small
velocities, the mentioned effects will be small and shall be neglected
in the following.

Note that an alternative way of determining the friction force in the
steady state is to time average the conservative plus the
non-conservative force that the upper wall exerts on the lubricant.
The observation that the work done by the conservative force is
essentially zero does not imply that its time-average must be zero.  A
formal derivation of the conclusions from this section is given in the
appendix.

\subsection{Effect of Instabilities}

\begin{figure}
  \includegraphics*[width=8.5cm]{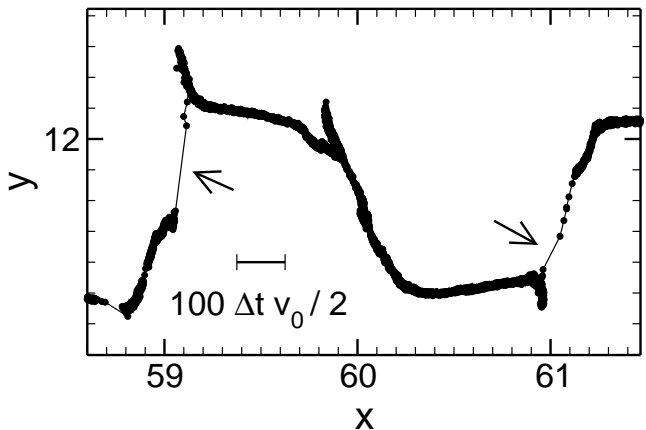}
\caption{ \label{fig:popping_example}
  Trajectory of a lubricant impurity in the $xy$-plane tagged between
  two incommensurate surfaces (at large load and small temperature).
  The relative velocities of the walls is $v_0 = 10^{-3}$.  The
  positions are plotted every $\Delta t = 0.5$.  The bar denotes 100
  times the average drift distance per time interval $\Delta t$.  The
  arrows indicate dynamical instabilities.}
\end{figure}

As discussed in the introduction, the externally imposed relative
motion of the confining walls may induce sudden, dynamic instabilities
or `pops' during which the particles' velocities greatly exceed both
their thermal velocities and the relative sliding velocity $v_0$ of
the walls.  This means that at a time $t + \delta t$ the atom does not
find a stable position in the ${\cal O}(v_0 \delta t)$ vicinity of the
old stable position at time $t$.  The continuous trajectory ends at
$t$ and the particle has to move to the next mechanically stable
position to resume its path.  The particle will then pop into the next
local potential minimum and for low sliding velocities, its peak
velocity $v_{\rm peak}$ will be solely determined by the energy
landscape and consequently $\lim_{v_0 \rightarrow 0^+} v_{\rm
  peak}/v_0$ diverges.  Its kinetic energy will be dissipated (e.g. by
phonons) and lead to friction.  This process will lead to a deviation
of the velocity distribution $P(v)$ from the thermal equilibrium
distribution $P_{\rm eq}(v)$ valid for $v_0 = 0$.
Fig.~\ref{fig:popping_example} shows such instabilities for a model
system that is described in detail in Sect.~\ref{sec:methods}.

The velocity distribution $P(v)$ and hence the friction force $F_{\rm
  k}$ can be calculated in principle, once the precise form of the
lubricant's interaction is known. Risken's book on the Fokker-Planck
equation~\cite{risken84} gives an excellent overview of methods that
allow one to treat models like ours, namely externally-driven systems
that are mainly deterministic but also contain a certain degree of
thermal noise.  An analytical approach remains difficult in our case,
due to the potential's complex time dependence.  Therefore a
different, phenomenological approach will be pursued.

An instability will invoke a trajectory during which potential energy
is abruptly converted into kinetic energy. The kinetic energy will
then be dissipated into the thermostat, i.e., the phonon bath of the
confining walls. After some time, which depends on the coupling
strength to the thermostat, the Maxwell-Boltzmann probability
distribution (PD) will be resumed, provided no new instability has
been invoked in the meantime. An instability will thus create a
typical velocity PD that will show up as a {\it tail} in the
Maxwell-Boltzmann PD.  Unless the two confining walls are identical
and perfectly aligned (thus commensurate), there is a class of
instabilities in which the energy lost during the `pop' shows a broad
distribution, see also Ref.~\onlinecite{daly02cm}. Every pop,
characterized for instance by the energy dissipated, will contribute
to $P(v)$ in its own way. We assume that the net sum $P_{\rm tail}$ of
all these individual tails shows exponential dependence on velocity,
thus
\begin{equation}
P_{\rm tail} \propto \exp{\left(-B\mid {\bf v} - {\bar{\bf v}} \mid\right)}\; ,
\label{eq:p_tail}
\end{equation}
where ${\bar{\bf v}}$ is the average velocity of the impurity under
consideration, typically ${\bf \bar{v}} = v_0 {\bf e}_x/2$, and $B$ is
a constant.  The motivation for this particular choice of $P_{\rm
  tail}$ partly stems from Jaynes' principle of information
theory~\cite{jaynes82}.  From it follows that the most likely
normalized PD on $[0, \infty)$ with given mean value about which we do
not have more knowledge is the exponential
distribution~\cite{HonerkampJaynes}, thus an exponential ansatz for
$P_{\rm tail}$ is plausible.  Further restraints lead to deviations
from an exponential form.  While these considerations are heuristic,
our choice of $P_{\rm tail}$ happens to be a quite accurate
description for the velocity PDs of impurities between 2D,
incommensurate surfaces.  This will be demonstrated later in the
result section.

At small sliding velocity $v_0$, the statistical weight of the tails
must increase linearly with velocity. Hence, the normalized PD
function for the $x$~component is given by
\begin{eqnarray}
P(v_x) & = & 
\sqrt{\frac{m}{2\pi k_{\rm B'} T}} \left(1 - \frac{2 A' v_0}{B'} \right)
{\rm e}^{-{m (v_x - \bar{v}_x) ^2}/{2 k_{\rm B} T}} \nonumber \\
& & + \, A' v_0 \rm{e}^{-B' |v_x - \bar{v}_x|} \;.
\label{eq:distribution}
\end{eqnarray}
Here $A'$, and $B'$ are phenomenological parameters that can (and
will) depend on the externally applied load $L$ that an impurity has
to counterbalance, damping $\gamma$, and other parameters.  However,
they should depend only weakly on temperature $T$ and sliding velocity
$v_0$ at small $T$ and small values of $v_0$.  This is because $A'
v_0$ is a measure for the rate of the fast processes (which should be
proportional to $v_0$ at small $v_0$), while $B'$ characterizes the
instability related velocity PD.  For 1D systems $B' = B$, the projection of a 2D exponential PD on one axis however leads to different $A'$ and $B'$~\cite{note_on_PD}.  Inserting Eq.~(\ref{eq:distribution}) into
Eqs.~(\ref{eq:dissipation}) and (\ref{eq:f_k}) and integrating over
$v_x$ yields the following friction force per impurity atom $F_{\rm
  k}/N_{\rm fl}$:
\begin{equation}
\frac{1}{N_{\rm fl}} F_{\rm k} =
4 \gamma m \frac{A'}{B'^3} - 2 \gamma \frac{A'}{B'} k_{\rm B} T \;,
\label{eq:fk_of_AB}
\end{equation}
from which the friction coefficient $\mu_{\rm k} = F_{\rm k} / (N_{\rm
  wall} L ) $ follows.  (Here, $L$ is the average load carried per
atom belonging to an outermost wall atom, thus $F_{\rm k}$ and $N_{\rm
  wall} L$ represent respectively the net friction force and the net
load.)  Of course, Eq.~(\ref{eq:fk_of_AB}) can only be valid as long
as Eqs.~(\ref{eq:p_tail}) and (\ref{eq:distribution}) give an accurate
description of the non-equilibrium velocity PD and provided that the
heat flow from the thermostat into the impurities is close to the
thermal equilibrium heat flow.  At extremely small $v_0$, two
arguments show that the assumption of exponential tails cannot
persist.  First, the energy $\Delta E_{\rm diss}$ that is dissipated
during a pop has an upper bound, which in turn implies an upper bound
for the peak velocity. Second, close to equilibrium, thermal noise is
sufficient to invoke (multiple) barrier crossing and recrossing.  The
ratio of sliding and noise-induced instabilities becomes small, which
in turn makes the non-equilibrium corrections be less significant.

Eq.~(\ref{eq:fk_of_AB}) is based on the one-dimensional distribution
functions $P(v_x)$. For two-dimensional interfaces and boundary
lubricants, the relevant PD is $P(v_{\|})$ with $v_{\|} =
\sqrt{v_x^2+v_y^2}$.  We want to note in passing that accumulating the
histogram $P(v_{\|})$ contains the same required information as
$P(v_x,v_y)$, however, it requires less data storage. This is why for
2-dimensional boundary lubricants, we monitor $P(v_{\|})$ instead of
$P(v_x,v_y)$.  Assuming rotational symmetry,
Eq.~(\ref{eq:distribution}) can be replaced with
\begin{equation}
P(v_{\|}) = 2 \pi v_{\|} A v_0 e^{-B v_{\|}} + 
\left( 1 - \frac{2\pi A v_0}{B^2} \right) P_{\rm eq}(v_{\|}),
\label{eq:2d_distribution}
\end{equation}
where $A$ and $B$ are phenomenological coefficients with similar
meanings as their counterparts $A'$ and $B'$ in the one-dimensional
description.

Inserting Eq.(\ref{eq:2d_distribution}) into
Eq.~(\ref{eq:dissipation}) yields:
\begin{equation}
\frac{1}{N_{\rm fl}} F_{\rm k} = 
12 \pi \gamma m \frac{A}{B^4} - 4 \pi \gamma \frac{A}{B^2} k_{\rm B} T
\label{eq:2d_friction}
\end{equation}

The parameters $A$ and $B$ will be obtained by fitting the PDs accumulated
during MD simulations. If such fits turn out to be good approximations
(and they do for incommensurate surfaces), then one has to ask into question
descriptions of frictional interfaces that are based on local effective
temperature (and local pressure). Local effective temperatures would imply
Gaussian rather than exponential velocity tails.  As we will argue later,
this difference might matter with regard to chemical reactivity
in a frictional interface.

Of course, the fits will never be perfect, and one has to address the
question whether one can obtain information experimentally on $A$ and $B$,
i.e., by measuring $F_k$ and the average kinetic energy of the lubricant
atoms. We will therefore compare the values of $F_k$ that are calculated
directly (by averaging the force on the top wall) with those that are obtained
indirectly with Eq.~\ref{eq:2d_friction} after $A$ and $B$ are
obtained through fits.

\section{Main Model and Methods}
\label{sec:methods}

In this paper, we analyze the trajectories of atoms and molecules
embedded between two walls in relative sliding motion by means of
molecular dynamics.  Different models with varying degree of
complexity are investigated ranging from rather simple, 1-dimensional
impurity models to 3-dimensional systems, in which the interaction
between the lubricant particles are taken into account.  In the latter
case, lubricant particles are not only simple atoms but may also
represent short polymers.  Here, we will only describe the methods
relevant to the full 3-dimensional simulations, as all other cases
only require 'dumbed-down' versions of that method or small
alterations thereof, such as suppressing the interaction between
lubricant atoms.

In our model, lubricant atoms interact with each other and with wall
atoms via a truncated Lennard Jones potential
\begin{equation}
  \label{eq:def_LJ_pot}
  V_{\rm LJ}(r) = \left\{ \begin{array}{r@{,\;}l}
      \! 4\epsilon  \left[ \left(\sigma/r\right)^{12}
        - \left(\sigma/r\right)^{6}\right] + C & r < r_{\rm c}\\
      0 & r \geq r_{\rm c}
    \end{array} \right. \;,
\end{equation}
where $r$ is the distance between two atoms.  $\epsilon$ defines the
energy scale and $\sigma$ the length scale of the system.  Both
quantities are set to unity.  A constant value $C$ is added to the
potential for inter-atomic distances smaller than the cut-off radius
$r_c$, which assures the continuity of the potential.  $r_{\rm c}$ was
chosen as the minimum of the LJ potential, $r_{\rm c} = r_{\rm
  min}=2^{1/6}$, unless otherwise noted.  This choice corresponds to a
purely repulsive interaction and can be justified by the observation
that at large pressures the essential behavior is caused by the
repulsion of the particles. The main effect of including the
attractive LJ contribution in the present context would be to add an
adhesive pressure.  Throughout the paper quantities are measured using
LJ units, such as time in units of $t_{\rm
  LJ}=(m\sigma^2/\epsilon)^{1/2}$ and forces in units of $\epsilon /
\sigma$. Atomic masses $m$ are also set to unity.

Both top and bottom wall lie in the $xy$ plane and consist of $N_{\rm
  wall}$ discrete atoms arranged in the hexagonal (1,1,1) plane
geometry of an fcc crystal.  The nearest neighbor spacing $d_{\rm nn}$
in the walls is $1.20914 \sigma$ unless noted otherwise.  This choice
of $d_{\rm nn}$ does not match with other length scales in the system.
Moreover, the relatively large value for $d_{\rm nn}$ enhances the
effect of surface corrugation.  In most simulations presented here,
the normal load is kept constant.  The load $L$ will be stated in
terms of normal load per atom in the upper wall, hence a unity normal
load corresponds to a pressure of about 0.79 in reduced quantities.

Commensurate wall geometries were realized by orienting the two walls
in parallel, whereas incommensurability was achieved by rotating the
upper solid surface by 90$^\circ$.  The use of periodic boundary
conditions in the wall plane required a slight distortion of the
perfect hexagonal geometry in order to obtain two quadratic walls.
Therefore, walls were not perfectly incommensurate anymore, but
quasi-incommensurate (as every setup realized with finite number
precision, strictly speaking). A wall unit cell consists of two atoms,
at positions $(0,0)$ and $(d_{\rm nn}/2, \sqrt3 \, d_{\rm nn}/2)$.  By
choosing the ratio of the wall unit cells in $\hat{x}$ and $\hat{y}$
close to the ideal value $\sqrt3$ this distortion was minimized.
We do not use other relative wall rotations in the full 3-dimensional
simulations, as it was found in the study of similar models that the
influence of the rotation angle is weak if it exceeds $\approx
5^{\circ}$~\cite{he01tl}.  We note that when two solids in an
experiment come in contact they will most likely be incommensurate, as
it would take utmost care to have two identical defect-free crystals
and orient them perfectly.  As detailed calculations show, elastic
deformations do not generally alter this argument provided the solids
are treated as 3-dimensional objects~\cite{muser03acp,shinjo93}.

While our analysis is focused on simple fluids, we include some work
on small chain molecules in order to study aspects of molecular
lubricants.  To this end, we used a well established bead-spring model
proposed by Kremer and Grest~\cite{KremerGrest1990}, which models
individual monomers as LJ particles while chain connectivity is
ensured by a FENE (finitely extensible nonlinear elastic) potential
given by
\begin{equation}
V_{\rm FENE} = 
-\frac12 k_{\rm ch}R_{\rm ch}^2\ln\left[1 - (r / R_{\rm ch})^2\right] \;,
\end{equation}
with $R_{\rm ch} = 1.5\sigma$ and $k_{\rm ch} = 30
\epsilon/\sigma^2$~\cite{KremerGrest1990}.  Typical values for
hydrocarbons are $\epsilon \approx 30 \text{meV}$, $\sigma \approx 0.5
\text{nm}$, resulting in a typical time scale of $t_{\rm LJ} \approx 3
\text{ps}$~\cite{KremerGrest1990}.

Simulations were done using a fifth order Gear-predictor corrector
algorithm (see e.g.~\cite{Haile1997}) with an integration step of $dt
= 0.005$. To maintain constant temperature, a stochastic Langevin
thermostat was employed~\cite{AllenTildesley}. It consists of ideal
white noise random forces and damping forces acting on all
thermostatted particles, which obey the fluctuation-dissipation
theorem~\cite{AllenTildesley}.  A damping constant $\gamma = 0.5$ was
used in all simulations. In the presence of instabilities, the precise
choice of $\gamma$ is usually quite irrelevant for friction forces at
small velocities~\cite{fisher85}, i.e., for the choice $\gamma_{\rm b}
=0.5$ and $\gamma_{\rm t} = 0$ we may expect similar friction forces
as for the perhaps more natural choice of $\gamma_{\rm b} =
\gamma_{\rm t} = 0.25$, if $v_0$ is sufficiently small.  However, in
either case, one must ensure that the random forces satisfy the
fluctuation dissipation theorem.
 
\section{Impurity Limit}
\label{sec:impurity}

\subsection{1D model systems}

Here, we want to discuss and extend those results from
Ref.~\onlinecite{muser02prl} that are relevant to this study.  In
Ref.~\onlinecite{muser02prl}, the following potential $V_{\rm t}$ for
the interaction between the impurity and the (one-dimensional) top
wall was employed
\begin{eqnarray}
V_{\rm t} & = & V_{{\rm t},0} \cos(2\pi (x-x_{\rm t})/b_{\rm t})  \nonumber \\
     & + &  V_{{\rm t},1} \cos(4\pi (x-x_{\rm t})/b_{\rm t}),
\label{eq:model_1d}
\end{eqnarray}
$b_{\rm t}$ being the lattice constant of the upper wall.  A similar
impurity-wall coupling is used to describe the interaction for the
bottom wall, however, the indices $t$ (for top) have to be replaced
with $b$ (for bottom).  This is a generalization of the interactions
suggested in Refs.~\onlinecite{rozman96prl} and
\onlinecite{rozman96pre} in that a non-zero first higher harmonic
$V_{{\rm t},1}$ is considered in Ref.~\cite{muser02prl}.  Moreover, the
lattice constant of the bottom wall $b_{\rm b}$ is allowed to differ
from that of the top wall $b_{\rm t}$.  We note in passing that
Refs.~\onlinecite{rozman96prl} and \onlinecite{rozman96pre} were
concerned with the interplay between external driving and embedded
system, while Ref.~\onlinecite{muser02prl} focused on the
constant-velocity friction in such systems.

In Ref.~\onlinecite{muser02prl}, it was found that the behavior of the
steady-state, low-velocity kinetic friction is surprisingly rich.  For
instance, athermal, zero-velocity friction turned out to vanish for
$V_{{\rm t},1} = V_{{\rm b},1} \le 0$.  In that case, athermal, small-velocity
friction can be described as a powerlaw $F_{\rm k} \propto v_0^\beta$
with a non-universal exponent $0 < \beta < 1$.  For $V_{{\rm t},1} = V_{{\rm b},1}
> 0$, $F_{\rm k}$ remains finite as $v_0$ approached zero.  The case
of $V_{{\rm t},1} = V_{{\rm b},1} = 0$ is particularly intriguing, as the minima
move at constant velocity $v_0/2$, then at some points in time (when
there is perfect deconstructive interference of $V_{\rm t}$ and
$V_{\rm b}$), the location of the potential energy minima make a
`phase jump' by a distance $b_{\rm t}/2$. This phase jump, however,
does not result in significant energy dissipation, as the location
from where the impurity stems and the location where the impurity ends
up are symmetrically equivalent.

\begin{figure}
  \includegraphics*[width=8.0cm]{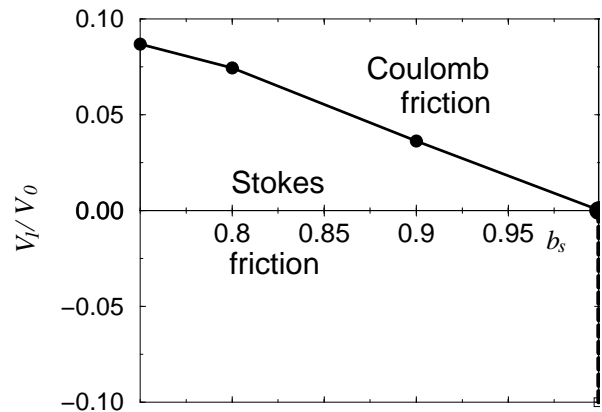}\hfill
\caption{ \label{fig:crit_points}
  Kinetic friction diagram for the impurity limit in 1 dimension.
  $V_1/V_0$ is the ratio of first-higher harmonic and fundamental
  harmonic in the lubricant-wall interaction. $b_{\rm s} = b_{\rm
    b}/b_{\rm t}$ is the ratio of the two lattice constants.  On the
  solid line and the dashed line, athermal kinetic friction is
  described by a power law $F_{\rm k}(v_0) \propto v_0^\beta$ with a
  non-universal exponent $0 < \beta < 1$.  In the Stokes regime below
  the solid line, $F_{\rm k} \propto v_0$.  In the Coulomb regime
  above the solid line, $F_{\rm k}(v_0 \to 0)$ remains finite. The
  solid line is an interpolation between the circles, which reflect
  points in the parameter space that have been explicitly
  investigated.  }
\end{figure}

For incommensurate walls ($b_{\rm t} \ne b_{\rm b}$), the behavior is
even richer.  If the first higher harmonic is not included, one wall
exerts a maximum force on the impurity and drags the impurity along.
As a consequence, $F_{\rm k}$ is linear in $v_0$, which we call Stokes
friction.  For one certain value of the first higher harmonic
$V_{{\rm t},1}^*$ (at a fixed ratio $b_{\rm s} = b_{\rm b}/b_{\rm t}$),
$F_{\rm k}$ can best be described as a power law in the limit of small
$v_0$.  For $V_{{\rm t},1} > V_{{\rm t},1}^*$, $F_{\rm k}$ remains finite in the
limit $v \to 0$, again provided thermal fluctuations are absent.
Fig.~\ref{fig:crit_points} shows a friction diagram for the 1D,
impurity-lubricant model. In order to yield a more complete picture
than that given in Ref.~\cite{muser02prl}, additional calculations
have been carried out, in order to determine $V_{{\rm t},1}^*$ for different
values of $b_{\rm s} = b_{\rm b}/b_{\rm t}$.

For a more detailed discussion of 
the effect of thermal fluctuations and the occurence of a Stokes
friction regime in one dimension and its potential applications,
we refer the reader to Ref.~\onlinecite{muser02prl}.

\subsection{2D model systems}
\label{sec:2D_systems}

\subsubsection{Model details}

We now allow the lubricant atoms to move within the $xy$-plane, but
motion normal to the interface in $z$ direction is still neglected.
Eq.~(\ref{eq:model_1d}) must then be replaced with a new model
potential.  As in other studies, we consider the symmetry of the
confining walls to be triangular, i.e., (111) surfaces of an fcc
crystal, for which the potential $V_{\rm t}$ between top wall and an
impurity can be written as:
\begin{equation}
V_{\rm t}({\bf x}, z) = \sum_{{\bf g}} \tilde{V}_{\rm t}({\bf g},z)
\exp\left\{i{\bf g} ({\bf x}-{\bf x}_{\rm t})\right\}.
\label{eq:model_2d}
\end{equation}
In Eq.~(\ref{eq:model_2d}), the sum goes over all {\it two}-dimensional 
reciprocal lattice vectors of the trigonal lattice ${\bf g}$,
${\bf x}$ is the position of the lubricant
particle in the $xy$ plane and ${\bf x}_{\rm t}$ the in-plane position
of the (top) wall.  $z$ denotes the (fixed) distance between (top)
wall and impurity.

The Fourier coefficients $\tilde{V}_{\rm t}({\bf
  g}, z)$ between chemically non-bonding species often {decay}
exponentially {with} $|{\bf g}|$ and increasing distance from the surface
$z$, thus $\tilde{V}_{\rm t}({\bf g},z)$ can be written as
\begin{equation}
\tilde{V}_{\rm t}({\bf g},z) = \tilde{V}_{\rm t}({\bf g},0) 
\exp\{- \alpha |{\bf g}| z\},
\label{eq:steele}
\end{equation}
where both parameters ${V}_{\rm t}({\bf g},0)$ and $\alpha$ depend on the
chemical nature of impurity atom and confining wall.  Potentials of
this form are known as Steele potentials~\cite{steele73}.  They have
proven to describe the potential energy landscape of atoms on
crystalline surfaces reasonably well~\cite{Patrykiejew2000}.  The
fundamental harmonic in this potential is related to the smallest
non-zero lattice vector ${\bf g} = (2\pi/\sqrt{3},0)$ and its five
symmetrically equivalent counterparts, which are obtained by rotating
${\bf g}$ successively by 60$^\circ$.  (Note that the distance between
'atoms' in the walls is set to unity, which differs from the choice
for the full 3D simulation model, see Sect.~\ref{sec:methods}.)  First
higher harmonics are related to reciprocal lattice vectors that are
the sum of a suitable pair of two different fundamental ${\bf g}$'s
and so on. The fundamental harmonic will be dominant at small loads,
however, as the external pressure increases (which makes $z$
decrease), the {\it relative} importance of higher harmonics will
increase due to Eq.~(\ref{eq:steele}).  The coupling between
impurities and bottom wall is similar to that between impurities and
top wall, however, the reciprocal vectors ${\bf g}$ are rotated an
angle $\theta$ with respect to the top wall's ${\bf g}$'s. Two walls
are called commensurate if $\theta$ is an integer multiple of
60$^\circ$.  An equivalent 2D model without higher harmonics, was used
recently by Daly et al.~\cite{daly02cm} for a study similar to that
presented here.

In the following, we will be concerned with an analysis of
mechanically stable position for the impurity atoms and their motion
as the walls slide against each other. The goal is to identify
situations, where the trajectory of a mechanically stable position
suddenly disappears, which would lead to a dynamical instability.
Such an analysis was given for 1D lubricants in in
Ref.~\cite{muser02prl}, see Fig.~1 in that paper, and also in
Ref.~\cite{daly02cm} for 2D systems.  For our analysis, the bottom
wall's lateral position ${\bf x}_{\rm b}$ is kept fixed, while ${\bf
  x}_{\rm t}$ is moved in small constant increments $d{\bf x}$ with
$|{d}{\bf x}| \equiv dx = 10^{-5}$ to $10^{-2}$. After identifying an initial
relative minimum in the impurity wall potential $V_{\rm i,w} = V_{\rm
  b} + V_{\rm t}$, a steepest descent algorithm (we used the Mathematica function FindMinimum[~]~\cite{MathematicaFindMinimum}) searches for the
closest minimum in $V_{\rm i,w}$ in the vicinity of the previous one.
If the distance $\Delta d$ between the location of the new and the old
minimum is greater than $\Delta d = 0.1$, we say that we identify a
{\it pop}.  While this choice of $\Delta d$ is somewhat arbitrary, we
ensured that our conclusions remained unaltered when $\Delta d$ was varied in reasonable bounds and $dx$ was further decreased.
 Since both methods employed in this study
(simulations and steepest descent) are exact and identical within 
controllable errors, their results mutually agree within these margins.

\subsubsection{Commensurate Walls}

Impurity atoms between commensurate walls only have a finite number of
non-equivalent minima in their potential energy landscape.  Once a
minimum is identified, symmetrically equivalent minima will exist at
periodically repeated positions that follow from the lattice of the
confining walls.  Various mechanically stable 'stacking' geometries
can be envisioned for our walls of trigonal symmetry, for instance
hexagonal close packed (hcp) and face cubic centered (fcc) type
configurations best characterized as respectively $ABA$ and $ABC$
layering structures.  The boundary lubricant reflects the middle
layer. While it does not correspond to an ideally crystalline layer,
the probability for a lubricant atom to sit at a certain position
would be indeed periodic, i.e. it would have a maximum in every single
$B$ position.

As the two walls are slid with respect to each other, the situation is
akin of the relative sliding of two commensurate, {\it
  one-dimensional} surfaces~\cite{muser02prl}.  The `trajectories' of
mechanically stable positions bifurcate and recombine at certain
relative, lateral displacements of the two solids, see Fig.~1a in
Ref.~\cite{muser02prl}.  This scenario invokes so-called {\it
  continuous} instabilities.  The peak velocities during continuous
instabilities tends to zero as $v_0$ tends to zero, however, this does
not happen linearly.  This will ultimately lead to the following
behavior if the walls are slid parallel to a symmetry axis and the top
wall is allowed to move freely in transverse direction: After every
half lattice constant moved, the system will convert from an fcc type
structure to an hcp type structure or vice versa.  This behavior is
also found in our 3D default system (see
Fig.~\ref{fig:part_pos+twall_pos}), in which impurities interact with
wall atoms through Lennard Jones potentials rather than through Steele
potentials.
The situation changes, when the top wall is not allowed to move in the
transverse direction, which was the choice in Ref.~\onlinecite{daly02cm}.
In that case, particles will occupy positions with high potential energy,
which leads to instabilities. In this study, however, we focus on the
case of zero transverse force.

As the mechanically stable positions of the embedded impurities show
no discontinuities, the kinetic friction force will tend to zero at
small $v_0$ even if thermal fluctuations are absent. From the
comparison to the 1D model systems, one would expect a powerlaw
behavior as in Eq.~(\ref{eq:athermal_fric}) with $F_{\rm k}(0) = 0$.
This behavior does not depend on the sliding direction.  It is also
observed for ratios of $V_{\rm t}({\bf g},z)/V_{\rm b}({\bf g},z)$
different from unity.

A central issue in the present paper is the question how robust the
property of the simple impurity model is as more complexity is added
to the model.  In the present case, one may argue that lubricant atoms
would be able to move in a correlated fashion up to a coverage of one
monolayer.  Above this coverage, the impurity model breaks down for
obvious reasons.

\subsubsection{Incommensurate Walls}

Impurity atoms between incommensurate walls have an infinite number of
inequivalent minima in their potential energy landscape for a given
relative wall displacement. This means that at a given moment in time,
it is impossible to find two different positions where the value of
the potential and all its derivatives are identical.  Yet, the number
of inequivalent trajectories of (meta)stable positions can be small,
because in most cases, they will all be identical up to temporal
shifts when the walls are in relative sliding motion.  See also the
discussion of the dynamics of the incommensurate Prandtl-Tomlinson
model by D. S. Fisher~\cite{fisher85}.

We analyze the instabilities by varying randomly the relative
orientation $\theta$ between the two walls as well as the sliding
direction $\phi$.  At this point, we are only concerned with the
occurrence of instabilities, rather than with the (average) amount of
energy dissipated during an instability.  Instabilities between
incommensurate walls are shown in Fig.~\ref{fig:popping_example}.

Unless $\theta$ is close to an integer multiple of 60$^\circ$, we find
that the number of instabilities depends only weakly on $\theta$ and
$\phi$. If we chose the fundamental harmonics $\tilde{V}({\bf g})$ of
both walls to be identical and the higher harmonics to be absent, then
we find on average one instability each time the upper wall has been
moved laterally with respect to the lower wall by a distance of
$200$~$d_{\rm nn}$.  Increasing the interaction strength for just one
wall does not change the behavior until the ratio $\tilde{V}_{\rm
  t}({\bf g})/\tilde{V}_{\rm b}({\bf g})$ or its inverse exceeds about
4.7. Above this threshold value, the metastable positions and hence
the particles follow the motion of just one wall and no instability
occurs.

Like Daly {\it et al.}~\cite{daly02cm}, we note that the instabilities
are possible due to transverse motion of the impurities, see also
Fig.~\ref{fig:popping_example}.  One of the issues Daly {\it et al.}
also discussed was the question above which value of $\tilde{V}_{\rm
  t}({\bf g})/\tilde{V}_{\rm b}({\bf g})$ the lubricant particle
remains pinned to the (top) wall. They reported a value of 4.5, while
we find a slightly higher value of 4.7, which essentially confirms the
prediction.  Furthermore, we also analyzed the effects of first higher
harmonics, which were neglected in Ref.~\onlinecite{daly02cm}.
Including the first higher harmonic ${\bf g}_1$ in addition to the
fundamental harmonics ${\bf g}_0$ increases the number of
instabilities, in particular for higher harmonics with positive sign.
Thus, the occurrence of instability remains a robust feature of
incommensurate walls.  For a ratio $V({\bf g}_1)/V({\bf g}_0)=0.1$,
the number of instabilities is increased by a factor of six. (At this
ratio the absolute value of the second harmonic would be in the order
of 0.01~$V({\bf g}_0)$, see Eq.~(\ref{eq:steele}), and will thus be
neglected.)  One may argue that the observed increase in pops is
related to an increase of incommensurability due to an additional
(small) length scale.

\subsection{3D Model}
\label{subsec:3d_model}

\subsubsection{Effect of commensurability on PDs}

\begin{figure}[hbtp]
  \includegraphics*[width=8.5cm]{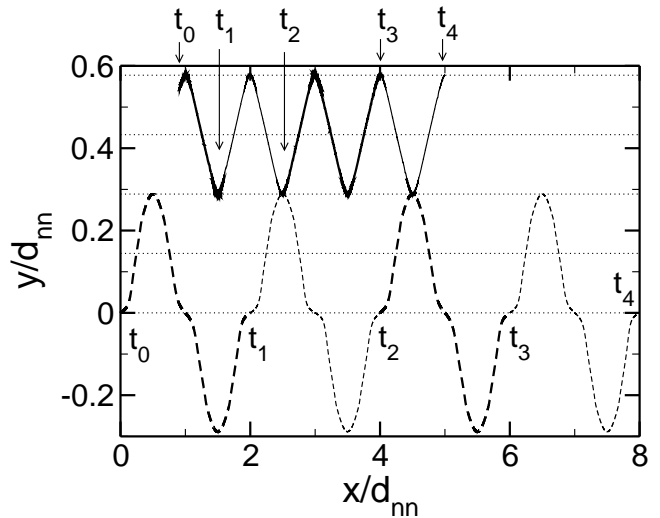}
\caption{ \label{fig:part_pos+twall_pos}
  Trajectory of a tagged particle (solid line) and of the upper wall
  (dashed line) for a commensurate system. The upper wall is moved
  parallel to $x$ at constant velocity. Horizontal lines are drawn at
  intervals $1/4\sqrt{3}$. For integer values of $x/d_{\rm nn}$ the
  configurations can be identified as hcp, for half-integer values as
  fcc configurations.  }
\end{figure}

We now turn to the analysis of the full three-dimensional model,
described in detail in Sect.~\ref{sec:methods}. Here, we also include
the interaction between the lubricant atoms.  However, as the coverage
is only a quarter layer, the results remain almost identical to the
ideal impurity limit.  Despite these changes with respect to
Section~\ref{sec:2D_systems}, all arguments discussed there remain
valid under the new conditions.  For instance,
Fig.~\ref{fig:part_pos+twall_pos} shows the expected dynamical
behavior of two {\it commensurate} walls separated by lubricant
impurities in sliding motion, i.e., an alteration of hcp and fcc type
configurations.  Most importantly, the trajectories of lubricant atoms
become continuous for commensurate walls. It is instructive to compare
Figs.~\ref{fig:popping_example} and \ref{fig:part_pos+twall_pos}.  For
completeness, we mention that the simulations in
Figs.~\ref{fig:popping_example} and \ref{fig:part_pos+twall_pos} were
both done at an external load of $L = 30$ per top wall atom, a thermal
energy of $T = 0.01$, and relative sliding velocity of $v_0 =
10^{-3}$.

\begin{figure}
  \includegraphics*[width=8.5cm]{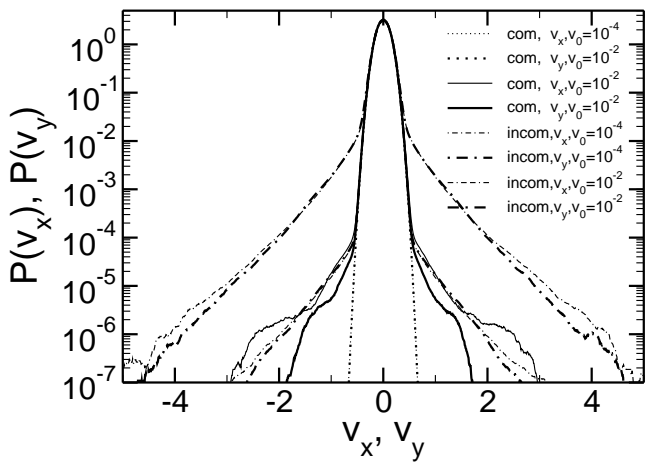}
\caption{ \label{fig:vel_dis_inc+com}
  Probability distribution (PD) of the fluid particles' x- and
  y-velocity components for shearing velocities $v_0 = 10^{-4}$ and
  $10^{-2}$ for incommensurate and commensurate wall orientations at
  $T=10^{-2}$ and $L=30$ for our standard system.  Around the central
  Maxwell-Boltzmann PD wide tails develop upon shearing.  The tails
  follow an exponential PD, see Eq.~(\ref{eq:distribution}), which
  have similar magnitude parallel and transversal to the sliding
  direction.  For commensurate walls, the tails are suppressed by two
  orders of magnitude at $v_0 = 10^{-2}$ and disappear completely for
  the lower shear velocity $v_0 = 10^{-4}$, when the PDs becomes
  almost indistinguishable from the Maxwell PD (not included).  }
\end{figure}

The different trajectories of the mechanically stable states result in
qualitatively different velocity distributions, even in the presence
of thermal fluctuations, which is shown in
Fig.~\ref{fig:vel_dis_inc+com}.  It can be seen that the velocity PDs
of impurities between incommensurate walls can indeed be described
with the non-equilibrium PD suggested in Eq.~(\ref{eq:distribution}).
It turns out that the PDs longitudinal ($x$) and transverse ($y$) to
the sliding direction ($x$) are almost identical. We note in passing
that the velocity PD normal to the interface ($z$ direction) is
affected much less than the in-plane PDs.

\subsubsection{Effect of sliding velocity and temperature on PDs}

As the relative sliding velocity between the walls is changed by a
factor of 100, the prefactor of the exponential tail scales with the
same factor, as suggested in Eq.~(\ref{eq:distribution}).  The
commensurate walls behave differently. First, the non-equilibrium
velocity distribution $P(v)$ deviates from equilibrium much less than
for incommensurate walls and it does not obey
Eq.~(\ref{eq:distribution}) as well.  More importantly, the tails of
$P(v)$ behave differently from those of incommensurate surfaces under
a change of sliding velocity.  This difference is due to the absence
of instabilities for the commensurate system. At $v_0 = 10^{-4}$, the
velocity PD for commensurate walls is almost identical to the
equilibrium Maxwell Boltzmann PD, while at the same $v_0$, the PDs for
incommensurate walls show distinct non-equilibrium tails. We employ a 
logarithmic scale for the PDs, because the tails can hardly be discerned 
on a linear scale.

\begin{figure}
  \includegraphics*[width=8.5cm]{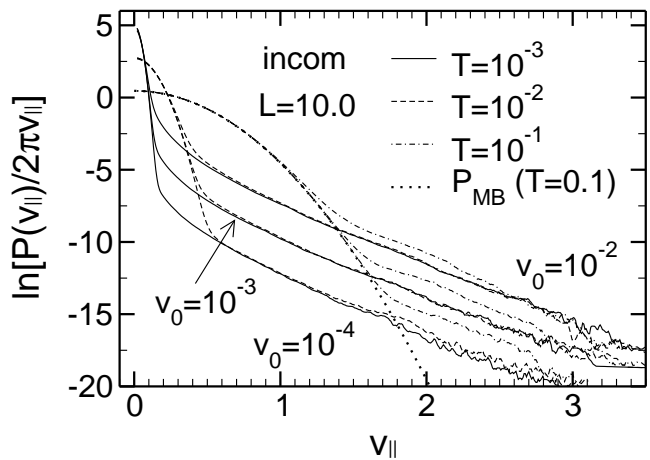}
\caption{ \label{fig:inc_detailcom}
  Probability distribution $\ln\left[P(v_\| )/2\pi v_\|\right]$ at
  load $L=10.0$ for temperatures $T=10^{-3}$, $T=10^{-2}$, $T=10^{-1}$
  and sliding speeds $v_0=10^{-3}$, $v_0=10^{-2}$, and $v_0=10^{-1}$.
  At low in-plane velocities $v_\|$ a thermal peak described by the
  Maxwell-Boltzmann PD (at $T=0.1$ exemplified by a thick dotted line)
  dominates, before the PD crosses over to exponentially distributed
  tails described in Eq.~(\protect\ref{eq:2d_distribution}).  The
  slope of the tails, $B$, is independent of both $T$ and $v_0$.  The
  prefactor of the tail distribution is proportional to $v_0$ and
  changes at large temperatures.  }
\end{figure}

Further examination of the distribution functions for incommensurate
surfaces as shown in Fig.~\ref{fig:inc_detailcom} reveals that the
coefficients $A$ and $B$ in Eq.~(\ref{eq:2d_distribution}) are
approximately constant for a wide range of velocities and
temperatures.  The parameters can be easily read off the graphs: The
slope of the tails equals $B$ and the exponential of the $y$-axis
intercept of a fitted line through the tails equals $Av_0$.  The data
for Fig.~\ref{fig:inc_detailcom} were produced with load $L=10.0$ for
temperatures $T=10^{-3} \ldots 10^{-1}$ and sliding speeds
$v_0=10^{-3} \ldots 10^{-1}$.

The present discussion is valid when the non-equilibrium tails are
clearly visible such as in Fig.~\ref{fig:inc_detailcom}.  It becomes
invalid when $v_0$ reaches extremely small values, i.e., when the
tails are starting to disappear under the central Maxwell-Boltzmann
peak.  Eq.~(\ref{eq:distribution}) then ceases to be a good
description of the PDs in that limit and Eqs.~(\ref{eq:distribution})
through (\ref{eq:2d_friction}) are no longer applicable.  However, the
equation describing the heat-flow balance between thermostat and
confined system, Eq.~(\ref{eq:dissipation}), is unaffected by this
argument and remains valid even in the limit $v_0 \to 0$.

\subsubsection{Effect of load on PDs}

\begin{figure}
  \includegraphics*[width=8.5cm]{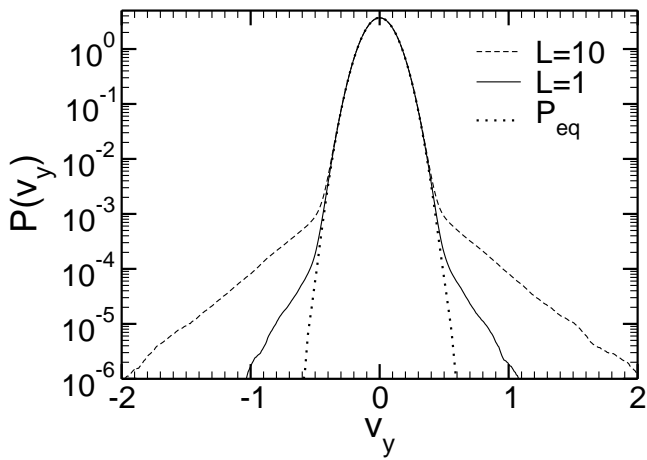}
\caption{ \label{fig:P_vy_T0.01_v0.001_L1+10A}
  Probability distribution $P(v_y)$ at $T = 0.01$, $v_0 = 10^{-3}$ and
  two different loads $L=1.0$ and $L = 10$. The thermal equilibrium
  distribution $P_{\rm eq}$ is inserted for comparison.  }
\end{figure}

The load dependence of the coefficients $A$ and $B$ was investigated
as well. We show the effect of load on the PDs for one of our model
systems exemplarily in Fig.~\ref{fig:P_vy_T0.01_v0.001_L1+10A}.  Many
similar calculations were done for other loads, coverages, and sliding
velocities with similar results for incommensurate surfaces.  In all
cases, we found that $A$ is roughly proportional to $L^{-0.8}$, while
$B$ is approximately proportional to $L^{-0.4}$.

From the normalization factor of the central equilibrium peak in
Eq.~(\ref{eq:2d_friction}), one may infer that the ratio $A/B^2$ is a
measure for the number of atoms far out of equilibrium and hence for
the number of invoked instabilities.  Given the proportionalities $A
\propto L^{-0.8}$ and $B \propto L^{-0.4}$, this number remains
constant when $L$ is increased.  Inserting the proportionalities $A
\propto L^{-0.8}$ and $B \propto L^{-0.4}$ into
Eq.~(\ref{eq:2d_friction}) results in a small deviation from
Amontons's law $F_{\rm k} \propto L$ at the microscopic level.

Potential differences scale with $L$ in lowest order, thus we obtain
for the energy dissipated in a pop $\Delta E_{\rm diss} \propto L
\propto v^2$.  Hence, for exact proportionality, the width of the
non-equilibrium tails was $\propto L^{0.5}$, respectively $B \propto
L^{-0.5}$, yielding Amonton's law.  This shows that $B \propto
L^{-\lambda} \;, \lambda \approx 0.5$ is to be expected, while the
precise value of the exponent $\lambda$ will depend on the specific
system potentials.

The deviation in our system is due to a shift of the relative
significance of lower- and higher-order harmonics.  This shift would
presumably be smaller if the repulsive forces were modeled with
(slightly more realistic) exponentially repulsive
forces~\cite{muser01prl}.

Fig.~\ref{fig:P_vy_T0.01_v0.001_L1+10A} reveals that the exponential
tails fall off less slowly when the pressure is increased. Thus, large
pressures in sliding contacts can dramatically increase the
probability of large velocities, even though the lubricant's average
kinetic energy $\langle T_{\rm kin}\rangle$ (or effective thermal
energy) may barely change.  This favours the occurrence of rare events
such as chemical bond breaking, as it becomes much more likely that a
bond is hit quasi-simultaneously by two high-velocity atoms. As the
non-equilibrium PDs fall off less slowly than the equilibrium PDs,
bond breaking will occur more frequently in non-equilibrium than in
equilibrium at a given thermal energy $\langle T_{\rm kin}\rangle$.
It will thus be difficult to assign a unique effective temperature
that reflects at the same time the reactivity of the molecules in the
junction and the energy contained in the vibrations.

\subsubsection{Comparison between calculated and measured friction coefficients}

The fit of curves equivalent to those shown in
Figs.~\ref{fig:vel_dis_inc+com} and \ref{fig:inc_detailcom} allows one
to estimate the kinetic friction force $F_{\rm k}$ with the help of
Eq.~(\ref{eq:2d_friction}).  This result can then be compared to the
friction force that is measured directly in the simulation.  It turns
out that such a comparison typically leads to an agreement within
approximately 25~\% accuracy, which can be improved by also taking
into account the effects of instabilities on the motion normal to the
surfaces.  The deviation between the `predicted' $F_{\rm k}$'s and the
directly measured $F_{\rm k}$'s is due to the fact that the tails are
not exactly exponential. This is particular important when the
temperature is large or $v_0$ extremely small.  If we accumulate the
correct $P(v)$'s in the simulation and use Eq.~(\ref{eq:dissipation})
to predict $F_{\rm k}$, the agreement between predicted and observed
kinetic friction is almost perfect, also when $v_0$ tends to zero.

\begin{figure}
  \includegraphics*[width=8.5cm]{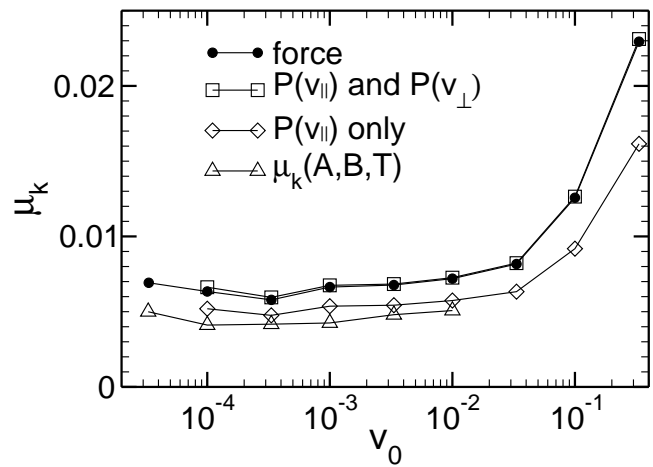}
\caption{ \label{fig:mu_d_comp_T0.001_L30}
  Comparison between the friction coefficient $\mu_{\rm k}$ as
  measured directly at the wall and calculated indirectly through the
  non-equilibrium velocity distributions $P(v)$. In two cases, the
  true distributions $P(v)$ were used. Taking into account both
  in-plane velocities $v_\|$ and velocities $v_\perp$ normal to the
  interface results in perfect agreement.  We also first fitted the
  PDs to Eq.~(\ref{eq:2d_distribution}), determined the coefficients
  $A$ and $B$ from the simulations and then calculated the kinetic
  friction force with Eq.~(\ref{eq:2d_friction}).  Quarter layer of
  lubricant, $T = 0.001$ and $L = 30$.  }
\end{figure}

Fig.~\ref{fig:mu_d_comp_T0.001_L30} shows the degree of agreement for
one particular model system. One can see that the kinetic friction
coefficients $\mu_{\rm k}$ as obtained from the full velocity PD, see
Eqs.~(\ref{eq:dissipation}) and (\ref{eq:f_k}) agree perfectly quite
well with the directly measured $\mu_{\rm k}$. Neglecting the
contribution of the motion normal to the surface results in an ${\cal
  O}(20\%)$ underestimation of the friction force. Estimating
$\mu_{\rm k}$ indirectly with the help of
Eqs.~(\ref{eq:2d_distribution}) and (\ref{eq:2d_friction}) leads to an
underestimation of about 25\%.

\subsubsection{Effect of temperature}
\label{sec:temperature}

It was shown by He and Robbins~\cite{he01tl} that the model system on
which this study is based yields logarithmic velocity corrections to
the friction force $F_{\rm k}$ for {\it incommensurate} surfaces,
provided the temperature is positive and the sliding velocity is not
too small, see also the discussion in Ref.~\cite{muser02prl}.  Our
simulation results of the $v_0$ corrections to $F_{\rm k}$ for
incommensurate surfaces are not shown explicitly, however, they
confirm the results by He and Robbins, The basic reason for a
logarithmic-type correction had already been recognized by
Prandtl~\cite{prandtl28}. Due to thermal fluctuations, the embedded
atoms can jump over local energy barriers and the instabilities will
be ignited prematurely. This reduces the necessary external force to
maintain sliding, because it does not need to move the embedded atom
all the way to the top of the energy barrier.

\begin{figure}
  \includegraphics*[width=8.5cm]{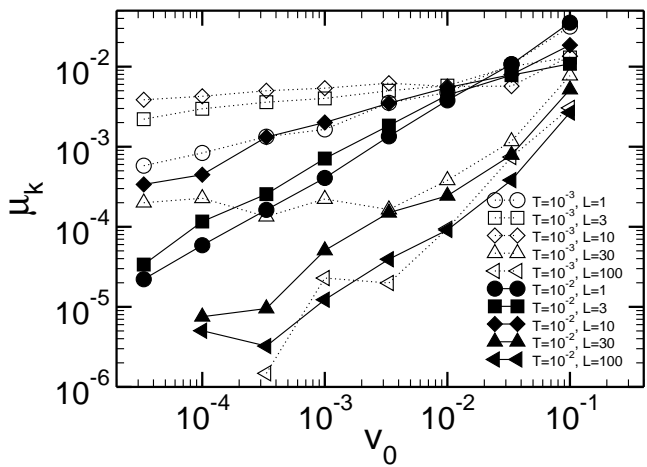}
\caption{ \label{fig:mu_d_2d_com_T0.001+0.01_0.25ml_allL}
  $\mu_{\rm k}$ of the commensurate standard system versus pulling
  velocity $v_0$ at different normal loads $L$ and temperatures $T$.
  Note the logarithmic scale for the $y$-axis.  In all cases $\mu_{\rm
    k}$ vanishes with a power law $v^\beta$ as $v \rightarrow 0$,
  except for $T=10^{-3}$, $L=30$ where a constant, small value seems
  to be reached.  }
\end{figure}

For {\it commensurate} surfaces, discontinuous instabilities are
absent and therefore the effect of thermal fluctuations must be
different.  This issue is investigated in
Fig.~\ref{fig:mu_d_2d_com_T0.001+0.01_0.25ml_allL}.  Due to the large
loads and the small temperatures employed, the linear-response regime
is not necessarily reached at the sliding velocities $v_0$ accessible
to the simulations, i.e., $v_0 = 10^{-5}$.  Therefore, we obtain
kinetic friction coefficients $\mu_{\rm k}$ that apparently vanish
according to
\begin{equation}
 \mu_{k} \stackrel{v_0\rightarrow 0}{\propto} v_0^\beta \;,
\end{equation}
with exponents $0.25 \lessapprox \beta \lessapprox 1$.

It is remarkable that a small change in temperature has a rather
strong effect on $F_{\rm k}$.  For the small load $L = 1$, the
exponent $\beta$ is approximately unity at temperature $T = 10^{-2}$
and one may argue that the corresponding $F_{\rm k}(v_0)$ reflects a
linear response curve.  As $T$ is lowered to $T = 10^{-3}$ , a
different exponent $\beta$ is obtained, reflecting non-equilibrium
behavior.  When the load is now increased by a factor of ten, the
energy barriers also increase approximately by a factor of ten.
Therefore the $F_{\rm k}(v_0)$ curves belonging to masses $L
\gtrapprox 10$ should be considered far from equilibrium, i.e
athermal. This would favour exponents $\beta$ less than unity.
However, this expectation is not true. Instead a Stokes-type friction
is observed. The almost linear relation of $F_{\rm k}$ and $v_0$ for
these largests loads ($L = 100$) may thus be an effect due to higher
harmonics in the lubricant-wall potential. As one can see in the 1D,
incommensurate systems, i.e., Fig.~\ref{fig:crit_points}, the
friction-velocity relationship can change qualitatively at certain
critical values of the higher-order harmonics.

\section{Beyond the impurity limit}
\label{sec:beyond}

So far, we have neglected the {\it direct} interactions between the
impurities or the coverages were small enough in order to render the
direct interactions negligible.  This approximation is reasonable when
the coverage is small and when the lubricant particles are simple
spherical units without inner degrees of freedom. When either
condition is violated, the energy landscape and hence the detailed
characteristics of the instabilities will change.  This in turn might
lead to a qualitative change in the tribological behavior of the
junction.  In this section, we will study the applicability, the
limitations and corrections of the impurity limit model that are due
to the interactions between lubricant atoms.

\subsection{Coverage effects}

When the lubricant coverage is close to or greater than one monolayer
and the junction is sheared, particles will have to move in a
correlated fashion. In order for one atom to jump to another
mechanically stable site, its neighbor has to jump as well, etc.  A
detailed description of the dynamics will be very complicated, i.e.,
it may involve sliding of correlated blocks along grain boundaries and
the formation of dislocation-type structures~\cite{persson00jcp}. Yet,
the argument persists that instabilities and sliding induced
deviations from the equilibrium velocity distribution function lead to
friction.

\begin{figure}
  \includegraphics*[width=8.5cm]{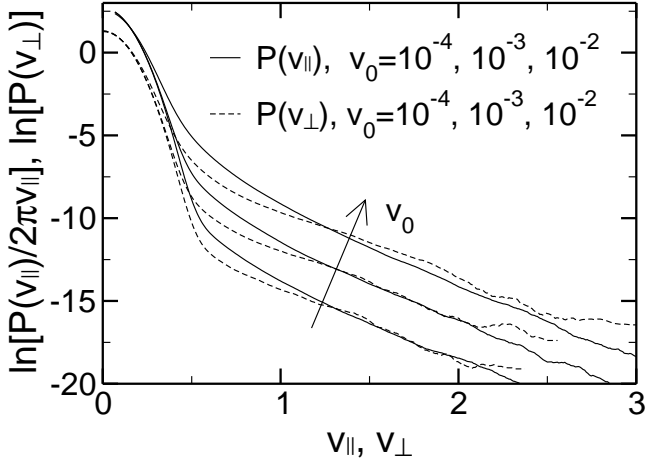}
\caption{ \label{fig:Pvz_Pvxy_incom_T0.01_L30_2.0ml}
  Distribution of the fluid particles velocity in plane, ($P(v_\|)$),
  and perpendicular to it, ($P(v_\bot)$), for an incommensurate system
  with two monolayers coverage at sliding velocities $v_0=10^{-4}$,
  $10^{-3}$, and $10^{-2}$.  The central Maxwell-Boltzmann parts are
  shifted because of the normalization $1/2\pi v_\|$ of $P(v_\|)$.  }
\end{figure}

Besides the correlated motion, some more details change when the
coverage is increased. For example, pops also occur in the direction
normal to the interface with a similar magnitude as parallel to the
interface.  This is reflected in the probability distributions $P(v)$
for the in-plane velocity $v_\|=\sqrt{v_x^2+v_y^2}$ and the normal
component $v_\perp = v_z$ of the fluid particles, see
Fig.~\ref{fig:Pvz_Pvxy_incom_T0.01_L30_2.0ml}.  The system under
consideration is incommensurate, the walls are separated by a double
layer and the externally imposed load per wall atom is $L = 30$.
Although the detailed dynamics of the lubricant atoms must be very
different from those in the impurity limit,
Eqs.~(\ref{eq:distribution}) and (\ref{eq:2d_distribution}) provide
again a reasonable description for respectively $P(v_\perp)$ and
$P(v_\|)$, i.e., a central Maxwell-Boltzmann peak and a
non-equilibrium exponential tail. Similar curves, which are not shown
explicitly, were obtained for a coverage up to 5 monolayers.

\begin{figure}
  \includegraphics*[width=8.5cm]{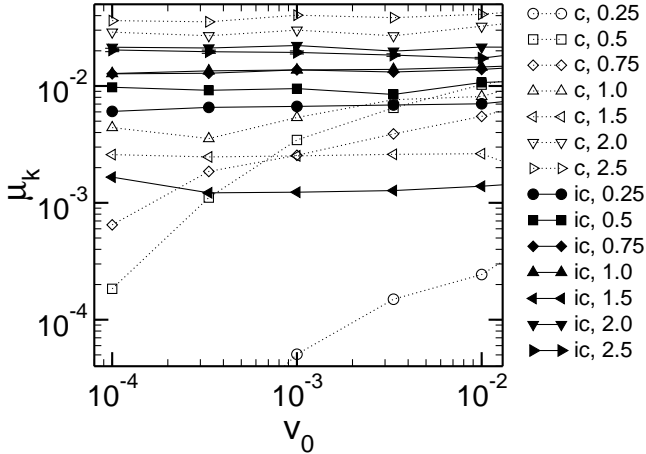}
\caption{ \label{fig:mu_d_2d_com+incom_T0.01_L30_Xml_vs_v0}
  Coverage dependence of the dynamic friction coefficient $\mu_{\rm
    k}$ of a system containing $0.25 \ldots 2.5$ monolayers of simple
  liquid at $T=10^{-2}$ and $L=30$. Commensurate systems (c) are
  denoted with open symbols, incommensurate walls (ic) with full
  symbols.  }
\end{figure}

As before, the kinetic friction force $F_{\rm k}$ is the integral over
the deviation of the $P(v)$'s from the Maxwell-Boltzmann distribution,
as stated in Eq.~(\ref{eq:dissipation}).  $F_{\rm k}$ is shown for
various coverages and sliding velocities in
Fig.~\ref{fig:mu_d_2d_com+incom_T0.01_L30_Xml_vs_v0}.  Both
commensurate and incommensurate systems are investigated and again
their behavior is strikingly different.

We start our discussion with the commensurate system. At a coverage of
$C = 0.25$, results are very close to the impurity limit.  $F_{\rm k}$
decays to zero with a powerlaw $v^\beta$ where the exponent $\beta$ is
less than one.  As the coverage is increased to $C = 0.5$ or even $C =
0.75$, $F_{\rm k}$ decreases considerably less quickly with decreasing
$v_0$ than in the impurity limit.  The behavior remains strikingly
different from Coulomb friction.  This changes when the coverage
reaches and exceeds one full monolayer.  For coverages beyond double
layers, the kinetic friction force even exceeds that of incommensurate
systems.  The prediction in Ref.~\onlinecite{muser02prl} that
commensurate systems should show smaller kinetic friction than
incommensurate system must thus be limited to extreme boundary
lubrication.  Above one monolayer lubrication, this trends seemingly
turns around.  Experiments suggest that commensurability leads to
enhanced friction between mica surfaces lubricated by a double layer
or more~\cite{ruths00}. Unfortunately, no study is known to the
authors in which a monolayer of lubrication or less was used between
two (smoothly) sliding commensurate walls.

At the smallest velocity investigated, $\mu_{\rm k}$ increases by a
factor greater than 200 for the {\it commensurate} case, when we
increase the coverage from $C = 0.5$ to $C = 2$.  The same change in
coverage for incommensurate surfaces only yields a factor of two.
Hence, incommensurate surfaces show much weaker coverage dependence
than commensurate interfaces.  Overall, there is relatively little
change of $F_{\rm k}$ with coverage for incommensurate walls with the
exception of $C = 1.5$.  Due to the large load employed, the
1.5~moloyers are squeezed into a single layer, which then essentially
acts like a solid.  This situation would not occur - or at least occur
only for a short period of time - if the lubricant could flow out of
the junction.  We conclude that the coverage dependence is weak for
incommensurate walls.

\subsection{Effects due to molecular bonds}

Most lubricant particles possess an inner structure.  Here we will
focus on the most simple generalization of the spherical molecules
considered so far, namely dimers, and hexamers (6-mers).  Dimers would
represent small linear molecules such as C$_2$H$_6$, while hexamers
are representative of short, linear alkane chains.  The dynamics of
the lubricant particles will change due to the additional internal
degrees of freedom. Alternatively, one may argue that the dynamics of
monomers is restricted because every monomer is constraint by at least
one chemically bonded neighbor.

\begin{figure}
  \includegraphics*[width=8.5cm]{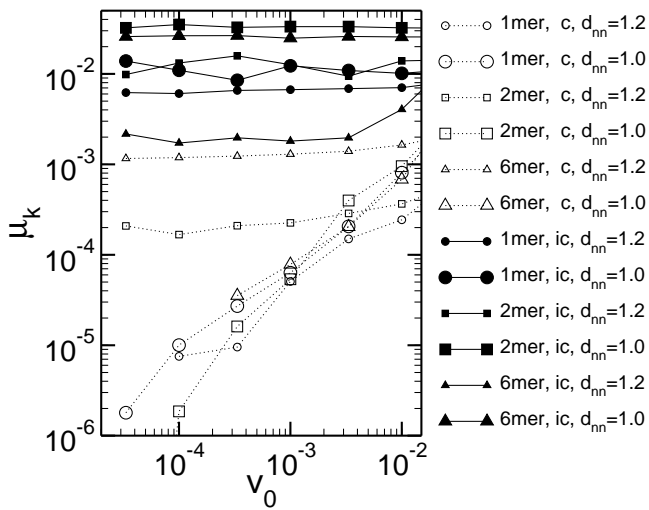}
\caption{ \label{fig:mu_d_2d_com+incom_1+2+6mer_0.25ml_dnn1.2+1.0_T0.01_L30_vs_v0}
  Dependence of the friction coefficient $\mu_{\rm k}$ on the wall
  lattice constant $d_{\rm nn}$ for a quarter monolayer of a dimer and
  a 6-mer with bond-length $d_{\rm mol} = 0.967$ for commensurate (c,
  open symbols) and incommensurate (ic, full symbols) orientation at
  small temperatures $T = 10^{-2}$ and large loads $L \approx 30$
  (adjusted to yield identical pressures).
}
\end{figure}

While monomers only have translational degrees of freedom, dimers also
have {\it rotational} degrees of freedom.  It is tempting to speculate
that 'rotational' instabilities can occur in addition to the
'translational' instabilities. Therefore, one might expect $F_{\rm k}$
to be larger for dimers than for monomers.  However, the rotational
and translation motion will not be independent of each other and the
coupling between them might reduce the effect of a 'translational'
instability. The question which effect dominates can only be answered
by analytical calculations or by molecular dynamics simulations.
Simulation results for the kinetic friction force in a
boundary-lubricated interface are shown in
Fig.~\ref{fig:mu_d_2d_com+incom_1+2+6mer_0.25ml_dnn1.2+1.0_T0.01_L30_vs_v0}.

In
Fig.~\ref{fig:mu_d_2d_com+incom_1+2+6mer_0.25ml_dnn1.2+1.0_T0.01_L30_vs_v0},
one can learn that the ratio $\rho = d_{\rm nn}/d_{\rm mol}$ of the
next neighbor spacing $d_{\rm nn}$ in the walls and the
intra-molecular bond length $d_{\rm mol}$ plays an important role,
particularly for commensurate surfaces. When the intra-molecular bond
length is close to the next-neighbor distance of wall atoms $d_{\rm
  nn}$, $F_{\rm k}$ disappears as a power law with sliding velocity
$v_0$. This means that the 'interference' effects between commensurate
walls persist and that no instabilities occur. Surprisingly, this is
even observed for hexamers. However, if $d_{\rm mol}$ differs from
$d_{\rm nn}$, instabilities also occur in boundary-lubricated systems,
even for commensurate walls. These instabilities are invoked through
the rotational degrees of freedom. While the misfit between $d_{\rm
  mol}$ and $d_{\rm nn}$ leads to Coulomb friction between
commensurate walls, its value of $F_{\rm k}$ remains small as compared
to the incommensurate case.

We now turn to the incommensurate walls.  Interestingly, the smoother
walls with $d_{\rm nn} =1.0$ produce higher kinetic friction than the
walls with $d_{\rm nn} = 1.2$, while the opposite is true for obvious
reasons in the case of static friction $F_{\rm s}$.  The reason for
the effect in $F_{\rm k}$ is that the reduced nearest neighbor spacing
leads to a higher rate of popping processes, as more atoms sliding
past each other at a given $v_0$ while the energy gain in the pops is
only slightly decreasing.  This effect can be verified by comparing
the distributions of the particle velocities.  The effect remains
stable for all degrees of polymerization.  The high friction of a
dimer is caused by the contribution of their fast rotations. This is
revealed by the distribution of the bonds' angular velocities.
Despite these trends, incommensurate systems prove again to be less
susceptible to quantitative changes in the parameters that determine
the details of the model than commensurate systems.

\section{Conclusions}
\label{sec:conclusions}

Kinetic friction requires the prevailance of instabilities (mechanical
hysteresis) in a system.  In the present paper, we have focused on
instabilities in the trajectories of particles confined between two
walls, which are sheared against each other.  When an instability is
reached, the particle does not find a local potential energy minimum
in its vicinity anymore and is thus forced to ``pop'' into the next
local minimum it sees.  At small sliding velocity $v_0$, this will
lead to a high velocity, which depends solely on the energy landscape.
The kinetic energy is gradually dissipated, resulting in a frictional
force.  We derived a relationship between the (non-equilibrium)
velocity distribution function $P(v)$ and the friction force $F_{\rm
  k}$. The characteristics of $P(v)$ and thus $F_{\rm k}$ depend only
weakly on coverage, sliding velocity, load and other parameters for
incommensurate surfaces.

In a generic setup, we first used two Steele potentials reflecting two
two-dimensional, triangular walls, which could be rotated with respect
to each other to achieve an incommensurate system.  We then computed
numerically the adiabatic trajectory of a test particle.  It was found
that instabilities were a robust feature of the incommensurate system.
Different off-symmetry wall rotations and inclusion of higher-order
contributions to the Steele potential as well asymmetric interaction
strengths of the walls did not alter the occurrence of the
instabilities, but only affected their frequency.

Including interactions between lubricant atoms does not change the
existence of instabilities and hence the presence of Coulomb friction
either.  In contrast, the commensurability of the walls allowed for
especially smooth trajectories of impurity atoms.  The trajectories
remain smooth when interactions between lubricant atoms are included
up to a coverage of one monolayer.  Above one monolayer, the lubricant
atoms do not move coherently any longer and instabilities are starting
to occur within the film. Kinetic friction rises dramatically as a
consequence.

We speculate that coherent motion similar to the one just described
may also be responsible for the behavior observed in a pioneering
quartz crystal oscillator study by Krim and
Chiarello~\cite{krim91JVTB}.  They found that the friction between a
{\it solid} monolayer and a smooth surface was much smaller than the
friction between a fluid monolayer and the same surface. The reverse
was reported for a rough surface.  Of course, the motion of a layer
adsorbed on a microbalance is different from that of a
''between''-sorbed layer, because in the first case, there is no
confining top wall and sliding-induced wall interference effects
cannot occur. This is an important qualitative difference, which
prevents us from making a direct comparison of our simulations with
the above-mentioned experiments.  One may yet argue that
commensurability can induce coherent motion of the film, be it
adsorbed or ''between''-sorbed. This suppresses erratic pops, which
ultimately lead to energy dissipation.  Thus, if one assumes that film
and smooth substrate were commensurate, the small values for the
kinetic friction force would not necessarily be in contradiction to
the supposedly large static friction force.  On the other hand for
rough, disordered surfaces, a solid monolayer would not be able to
move coherently, which would be consistent with its large friction as
compared to a fluid layer.

We turn back to the discussion of the non-equilibrium velocity
distributions. For incommensurate walls, the distribution consists of
a central peak, which is essentially identical to the equilibrium
velocity distribution, and of an additional non-equilibrium tails.
These tails fall off only exponentially with $v$, which is slower than
the exponential decay with $v^2$ in equilibrium systems. This
observation is rather generic for incommensurate systems and
independent of the lubricant coverage.  As the real velocity
distribution function is qualitatively different from Gaussians, it
seems futile to describe the interface in terms of an effective
temperature. We argued that given a specific kinetic energy associated
with a lubricant (which could be used to define an effective
temperature), the non-equilibrium system would be more likely to
invoke chemical bond breaking or other chemical reactions.

Overall, the impurity model provides a good description of the typical
characteristics of a boundary-lubricated system.  However, it is
essential to study two-dimensional interfaces and incommensurate
surfaces. One-dimensional and/or commensurate surfaces lead to
untypical behavior, i.e., rather large sensitivity of the friction
force with respect to small changes in the model (details of
interaction potential) or in the external parameters (sliding
velocity, load, temperature, etc.). This is unfortunate, because
incommensurate walls are much more common than commensurate walls,
which leaves us with fewer possibilities to control friction.

A surprising result of our study for incommensurate walls is that
increasing the atomic scale roughness of the walls may actually
sometimes reduce the kinetic friction force.

It would be interesting to compare our predictions concerning the
velocity distributions to experimental data. While scattering data
from small, confined volumes is certainly notoriously difficult to
obtain, recent advances have been made.  Using fluorescence
correlation spectroscopy, Mukhopadhyay et. al.~\cite{mukhopadhyay02}
measured translational diffusion in molecularly thin liquids confined
within a surface forces apparatus. In the future, it might be possible
to extend these studies to sliding situations so that velocity
distributions can be measured.


\begin{acknowledgments}
\label{sec:acknowledgments}
M.A. is indebted to \mbox{J. Baschnagel} and \mbox{M. Fuchs} for
numerous discussions and acknowledges support by the German Academic
Exchange Service (DAAD, ``Hochschulsonderprogramm III von Bund und
L\"andern''), grant no.\ D/00/07994.  We are also grateful to the
Bundesministerium f\"ur Bildung und Forschung (BMBF) for partial
support under grants no.\ D.I.P.\ 352-101 and 03N6500.
\end{acknowledgments}

\bibliography{instab}

\begin{thebibliography}{10}
\bibitem{note_on_PD}
\bibinfo{note}{The projection $P(v_x) = \int P(v_x,v_y) dv_y = \int \exp[-B
  \sqrt{v_x^2 + v_y^2}] dv_y$ is in general only approximatively exponential.
  However, for sufficiently large $v_x$, this approximation is very accurate
  and $B'$ with $\exp[-B' v_x] \simeq P(v_x)$ is close to $B$.}
\end{thebibliography}

\appendix
\section{Relation between velocity distribution and friction}

Consider a system in steady-state equilibrium with the following
underlying equation of motion
\begin{equation}
m\ddot{\bf x} + m \gamma \dot{\bf x} = {\bf F}_{\rm b}({\bf x})
+ {\bf F}_{\rm t}({\bf x}-{\bf v}_0t) + {\bf \Gamma}(t).
\label{eq:equation_of_motion}
\end{equation}
Here, we chose the same terminology as in Sect.~\ref{sec:theory},
i.e., ${\bf F}_{\rm b}({\bf x})$ denotes the force of the bottom wall
on an impurity atom located at position ${\bf x}$ and ${\bf v}_0$ is
the velocity of the upper wall with respect to the substrate.  We
multiply Eq.~(\ref{eq:equation_of_motion}) with $\dot{\bf x}$ and
average over a long time interval $\tau$.  We then interpret the
resulting individual terms. They can be associated with the (average)
power dissipated within the system or the (average) power put into the
system.  First, the average change of kinetic energy with time equals
zero, namely
\begin{eqnarray}
\frac{1}{\tau} \int_0^{\tau} dt \, m\ddot{\bf x} \dot{\bf x} & = &
\frac{1}{\tau} \int_0^{\tau} dt \, \frac{d}{dt} T_{\rm kin}  \nonumber\\ & = &
\frac{1}{\tau} [T_{\rm kin}(t=\tau) - T_{\rm kin}(t=0)] \nonumber\\ & \to &
0 \;\; {\rm for } \;\; \tau \to \infty
\end{eqnarray}
The second term is proportional to the time-averaged kinetic energy of
the system with respect to the lower wall:
\begin{eqnarray}
\frac{1}{\tau} \int_0^{\tau} dt \, m \gamma \dot{\bf x} \dot{\bf x} & = &
\gamma m \langle \dot{\bf x}^2 \rangle \nonumber\\ & = &
2 \gamma \langle T_{\rm kin} \rangle,
\label{eq:T_kin_ave}
\end{eqnarray}
$\langle T_{\rm kin} \rangle$ being the time-averaged or
ensemble-averaged (steady-state) kinetic energy of an impurity.
Thermostating also parallel to the top wall, e.g.\ by choosing
$\gamma = \gamma_{\rm t} = \gamma_{\rm b}$, requires a trivial modification of
the reference system.  The next term is the average work of the bottom
wall on the impurity
\begin{eqnarray}
\frac{1}{\tau} \int_0^{\tau} dt \, \dot{\bf x} {\bf F}_{\rm b}({\bf x})
& = & \frac{1}{\tau} \int_0^{\tau} dt \, \left(-\frac{d}{dt} V_{\rm b}({\bf x}) \right)
\nonumber\\
& = &
\frac{1}{\tau} \left\{ V_{\rm b}[{\bf x}(\tau)] - V_{\rm b}[{\bf x}(0)]\right\}
\nonumber\\ & \to & 0,
\label{eq:work_bottom}
\end{eqnarray}
which must vanish in any steady-state system. Of course, if the model
was generalized such that (steady state) wear would occur, then the
contribution discussed in Eq.~(\ref{eq:work_bottom}) would indeed
remain finite.

For the discussion of the next term in
Eq.~(\ref{eq:equation_of_motion}), it is necessary to keep in mind
that
\begin{equation}
\frac{d}{dt} V_{\rm t}({\bf x}-v_0t) = -F_{\rm t}({\bf x}-v_0t) \left[ \dot{\bf x}-v_0
\right].
\end{equation}
This and the same considerations invoked for
Eq.~(\ref{eq:work_bottom}) yield
\begin{equation}
\frac{1}{\tau} \int_0^{\tau} dt \, \dot{\bf x} {\bf F}_{\rm t}({\bf x}-{\bf v}_0 t)
= \langle {\bf F}_{\rm t} \rangle {\bf v}_0 \;,
\end{equation}
where $\langle {\bf F}_{\rm t} \rangle$ is the time- or
ensemble-averaged force that the top wall exerts on an impurity. This
force or depending on the definition its projection onto the sliding
direction can be associated with the kinetic friction force $F_{\rm
  k}$.

The contribution due to the random force ${\bf \Gamma}(t)$ is the most
difficult contribution to calculate. However, if the system is close
to local equilibrium for most of the time, then the expectation value
of ${\bf \Gamma}(t)\dot{\bf x}$ can be expected to be close to the value of
this expression in thermal equilibrium. In equilibrium, it must
compensate the expression discussed in Eq.~(\ref{eq:T_kin_ave}), hence
\begin{equation}
\frac{1}{\tau} \int_0^{\tau} dt \, \dot{\bf x} {\bf \Gamma}(t) \approx 
2 \gamma \langle T_{\rm kin} \rangle_{\rm eq} \;,
\end{equation}
where $\langle T_{\rm kin} \rangle_{\rm eq}$ denotes the average
kinetic energy in thermal equilibrium.

Assembling all necessary terms, yields
\begin{equation}
2\gamma (\langle T_{\rm kin} \rangle - \langle T_{\rm kin} \rangle_{\rm eq})
= F_{\rm k} v_0 \;.
\label{eq:final}
\end{equation}
 Note that Eq.~(\ref{eq:final}) is equivalent to Eqs.~(\ref{eq:dissipation}) and (\ref{eq:f_k}).

\end{document}